\documentclass[12pt,showpacs,amsmath,amssymb,]{revtex4}
\usepackage{graphics}
\usepackage{amsfonts}
\usepackage{amsmath}
\usepackage{epsfig}
\usepackage{epsf}
\usepackage{graphicx}
\usepackage{dcolumn}
\usepackage{bm}
\usepackage{color}
\usepackage{array}

\newcommand {\sla}[1]{ #1 \!\!\!/}

\begin{document}
\title{ Meson exchange effects in elastic $ep$ scattering at loop level and the electromagnetic form factors of the proton }

\author{
Hong-Yu Chen$^{1}$ , Hai-Qing Zhou$^{1,2}$\protect\footnotemark[1] \protect\footnotetext[1]{E-mail: zhouhq@seu.edu.cn}\\
$^1$Department of Physics,
Southeast University, NanJing 211189, China\\
$^2$State Key Laboratory of Theoretical Physics, Institute of Theoretical Physics,\\
Chinese Academy of Sciences, Beijing\ 100190,\ P. R. China }

\date{\today}

\begin{abstract}

A new form of two-photon exchange(TPE) effect is studied to explain the
discrepancy between unpolarized and polarized experimental data in elastic $ep$ scattering.
The mechanism is based on a simple idea that apart from the usual TPE effects from box and crossed-box diagrams, the mesons may also be exchanged in elastic $ep$ scattering by two-photon coupling at loop level.
The detailed study shows such contributions to reduced unpolarized cross section ($\sigma_{un}$) and polarized observables ($P_t,P_l$)
at fixed $Q^2$ are only dependent on proton's electromagnetic form factors $G_{E,M}$ and a new unknown universal parameter $g$. After combining this contribution with the usual TPE contributions from box and crossed-box diagrams, the ratio $\mu_pG_E/G_M$ extracted from the recent precise unpolarized and polarized experimental data can be described consistently.
\end{abstract}

%\textbf{Key words:} Two-Photon Exchange, Form Factor
\maketitle
%%%%%%%%%%%%%%%%%%%%%%%%%%%%%%%%%%%%%%%%%%%%%%%%%%%%%%%%
\section{Introduction}

As the basic constituent of our world and most elemental bound states of strong interaction, the proton plays an important role in the physics. Up to now, our knowledge on the structure of proton has still been poor, for example, how big is the proton\cite{proton-size}, how large are the electromagnetic form factors $G_{E,M}$ of the proton\cite{Ex-polarized,Ex-polarized-Meziane-2011,Ex-Rosenbluth-1994,Ex-Rosenbluth-2006}. Since the first measurement of $R=\mu_p G_E/G_M$ by the polarization transfer (PT) method\cite{Ex-polarized}, it becomes a serious problem for theoretical physicists to explain the large discrepancy of extracted $R$ between the PT method and Rosenbluth or longitudinal-transverse (LT) method\cite{Ex-Rosenbluth-1994,Ex-Rosenbluth-2006}.

\begin{figure}[htbp]
\center{\epsfxsize 3.0 truein\epsfbox{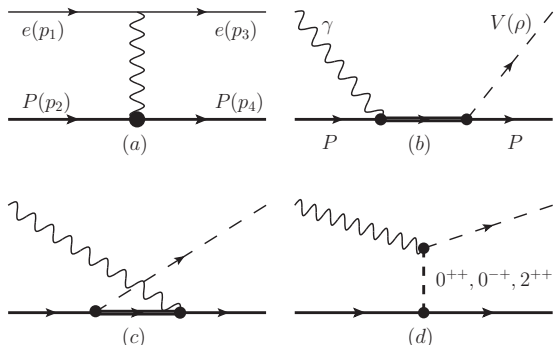}}
\caption{(a)The Born diagram in elastic $ep$ scattering. (b,c,d) The $s$,$u$,$t$ channels in photoproduction of vector meson, the similar diagrams in VCS are not shown.}
\label{Born-photo-production}
\end{figure}

In the Born approximation, the elastic $ep$ scattering is described by one-photon exchange (OPE) shown in Fig. \ref{Born-photo-production}(a).
By this approximation, the reduced unpolarized cross section is expressed as
\begin{eqnarray}\label{OPE-sigma}
\sigma_{un,th}^{1\gamma} \equiv \left. {\frac{d\sigma^{(un)} }{d\Omega}} \right|_{lab} \frac{\varepsilon (1 + \tau )}{\tau \sigma _{ns} } ={ G_M^2 + \frac{\varepsilon}{\tau} G_E^2 },
\end{eqnarray}
and the polarized observables $P_t,P_l$ are expressed as
\begin{eqnarray}\label{OPE-PTPL}
P_{t,th}^{1\gamma} &=& -\frac{1}{\sigma_{un,th}^{1\gamma}}\sqrt {2\varepsilon(1 - \varepsilon)/\tau} G_M G_E, \\ \nonumber
P_{l,th}^{1\gamma} &=& \frac{1}{\sigma_{un,th}^{1\gamma}} \sqrt {(1 + \varepsilon )(1 - \varepsilon )} G_M^2, \\ \nonumber
R_{PT,th}^{1\gamma} &\equiv& -\mu_p\sqrt{\frac{\tau(1+\epsilon)}{2\epsilon}}\frac{P_{t,th}^{1\gamma}}{P_{l,th}^{1\gamma}}=\mu_p \frac{G_E}{G_M},
\end{eqnarray}
with $\sigma_{ns}=\frac{\alpha^2cos^2(\theta_e/2)}{4E^2sin^4(\theta_2/2)}\frac{E'}{E}$, $\tau=Q^2/4M_N^2,Q^2=-q^2,q=p_1-p_3,\epsilon=[1+2(1+\tau tan^2\theta_e/2)]^{-1}$, $M_N$ the mass of proton, $\alpha$ the fine structure constant, $\theta_e$ the scattering angle of electron, $E$ and $E'$ the energies of initial and final electrons in the laboratory frame, respectively. The detail of the physical meaning of $P_{t,l}$ can be seen in the literature, for example, \cite{Ex-polarized}.

Experimentally, the LT method extracts $R$ from the $\epsilon$ dependence of an experimental unpolarized cross section at fixed $Q^2$ by Eq.(\ref{OPE-sigma}) and the PT method extracts $R$ from the experimental ratio $P_t/P_l$ at fixed $Q^2$ and $\epsilon$ by Eq.(\ref{OPE-PTPL}). In the following we name such extracted $R$s as $R_{LT,Ex}^{1\gamma}$ and $R_{PT,Ex}^{1\gamma}$, respectively. The current precise experimental measurements\cite{Ex-polarized,Ex-Rosenbluth-2006} show that $R_{LT,Ex}^{1\gamma}$ are much larger than $R_{PT,Ex}^{1\gamma}$  when $Q^2>$2GeV$^2$.

In the literature, two-photon exchange (TPE) effects are suggested to explain such a discrepancy \cite{TPE-review}. Many model dependent methods are studied to estimate the TPE corrections such as the simple hadronic model \cite{TPE-hadronic-model}, GPDs method \cite{TPE-GPDs}, dispersion relation method \cite{TPE-dispersion-relation}, pQCD \cite{TPE-pQCD}, and SCET \cite{TPE-SCEF}.
These model dependent calculations gave similar TPE corrections to $R_{LT,Ex}^{1\gamma}$, and it is usually concluded that the discrepancy is able to be explained by TPE corrections \cite{TPE-hadronic-model,Arrinton2007}. But the recent polarized experimental data \cite{Ex-polarized-Meziane-2011} show very different properties of TPE corrections to $R_{PT,Ex}^{1\gamma}$ with that predicted by these theoretical models. For example, the experimental data showed that the TPE corrections to $R_{PT,Ex}^{1\gamma}$ are almost a constant at $\epsilon=(0.152,0.635,0.785)$ when $Q^2=2.49$ GeV$^2$ \cite{Ex-polarized-Meziane-2011}, while the theoretical estimations of TPE corrections are large and positive at small $\epsilon$ by the hadronic model and dispersion relation method \cite{TPE-hadronic-model,TPE-dispersion-relation}, and are large and negative at small $\epsilon$ by the GPDs method and pQCD method \cite{TPE-GPDs,TPE-pQCD}. This situation shows that we are still far away from the accurate understanding of experimental data in elastic $ep$ scattering. And a further careful study of TPE corrections or similar effects are strongly called for.

\begin{figure}[htbp]
\center{\epsfxsize 3.0 truein\epsfbox{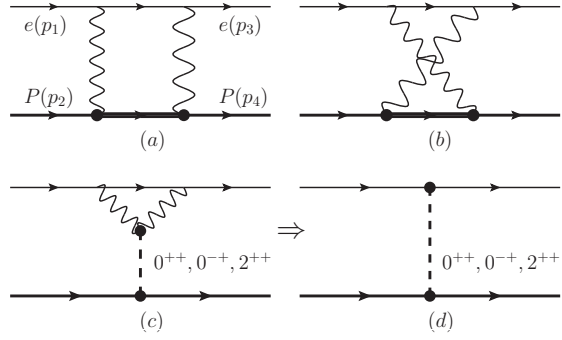}}
\caption{TPE contributions in $ep$ scattering. (a) box diagram; (b) crossed-box diagram; (c) meson-exchange diagram by two-photon coupling; (d) effective direct meson-exchange diagram.}
\label{ep-scattering}
\end{figure}

In this work, we consider a new form of TPE effect in elastic $ep$ scattering. The main idea is from the theoretical estimations of virtual Compton scattering(VCS) and photoproduction of the vector meson. For these two processes, the contributions from the $s$, $u$, and $t$-channels shown in Figs. \ref{Born-photo-production}(b,c,d) are usually all included in the effective models \cite{meson-exchange,Regge-meson-exchange}. When considering the radiative corrections in elastic $ep$ scattering, it is natural that the corresponding similar contributions shown as Figs. \ref{ep-scattering}(a,b,c) will give contributions, where only the permitted spin 0 and 2 mesons are included in the $t$ channel. Figures \ref{ep-scattering}(a,b) are just the usual box and crossed-box diagrams studied in \cite{TPE-hadronic-model}, while the contribution from Fig.\ref{ep-scattering}(c) is usually ignored in the literature. In Sec. II, at first we rewrite the contribution from Fig.\ref{ep-scattering}(c) in a simple and general form by the effective interactions, and then present the expressions for the reduced unpolarized cross section and polarized observables after including this contribution. In Sec. III, we present our numerical analysis on the recent experimental data, the TPE corrections to the extracted $R$ by LT and PT methods, and the TPE contributions to the ratio between unpolarized cross sections of elastic $e^+p$ and $e^-p$ scattering.

\section{Basic Formula}

The formal gauge invariant couplings of $M\gamma\gamma$ in Fig.\ref{ep-scattering}(c) can be written down similarly with those in \cite{meson-exchange,Regge-meson-exchange}, while in the case of Fig. \ref{ep-scattering}(c), the two virtual photons are in the loop and their momentums are not limited by any conditions except their sum. This is different with the usual VCS case where the coupling constants are taken as constants or multiplied by some special form factors in a special kinematic region. To avoid the uncertainty from the momentum dependent coupling constants and describe the effect in a reliable and universal form, we rewrite the contributions from Fig.\ref{ep-scattering}(c) in a general effective direct meson-exchange form shown as Fig.\ref{ep-scattering}(d) where all the momentum dependence of $M\gamma\gamma$ couplings and their integrations are absorbed into the effective couplings between electron and mesons, and the new effective couplings now are only dependent on $Q^2$. The most general form of the effective interactions for $0^{++}, 0^{-+},2^{++}$ mesons can be written as
\begin{eqnarray}\label{effective-interactions}
\Gamma_{See}&=&-ig_{See},~~~~\Gamma_{Spp}=-ig_{Spp}, \\ \nonumber
\Gamma_{Pee}&=&g_{Pee,1}\gamma_{5}-ig_{Pee,2}\gamma_{5}(\sla{p}_f-\sla{p}_i), \\ \nonumber
\Gamma_{Tee,\mu\nu}&=&g_{Tee,1}(p_f+p_i)_{\mu}\gamma_{\nu}-ig_{Tee,2}g_{\mu\nu}, \\ \nonumber
\Gamma_{Ppp}&=&g_{Ppp,1}\gamma_{5}-ig_{Ppp,2}\gamma_{5}(\sla{p}_f-\sla{p}_i),\\ \nonumber
\Gamma_{Tpp,\mu\nu}&=&g_{Tpp,1}(p_f+p_i)_{\mu}\gamma_{\nu}-ig_{Tpp,2}g_{\mu\nu},
\end{eqnarray}
where $S,P,T$ refer to the scalar, pseudoscalar, and tensor meson, $p_i,p_f$ refer to the initial and final momentums of electron and proton, and all the couplings $g_i$ are only functions of $Q^2$.
The propagators of exchanged mesons are taken as the Regge form \cite{Regge-meson-exchange}
\begin{eqnarray}\label{propagator}
S_{S,P}(q)&=&\mathcal{P}_{S,P}(q), \\  \nonumber
S^{\mu\nu;\rho\omega}_{T}(q) &=&\Pi^{\mu\nu;\rho\omega}(q)\mathcal{P}_{T}(q),
\end{eqnarray}
where $\Pi^{\mu\nu;\rho\omega}(q)=\frac{1}{2}(\eta^{\mu\rho}\eta^{\nu\omega}+\eta^{\mu\omega}\eta^{\nu\rho})-\frac{1}{3}\eta^{\mu\nu}\eta^{\rho\omega}$,
$\eta^{\mu\nu}=-g^{\mu\nu}+q^\mu q^\nu/m_T^2$ and
\begin{eqnarray}\label{Regge}
\mathcal{P}_{X}
&=&\frac{\pi\alpha'_{X}}{\Gamma[\alpha_{X}(t)-J_{X}+1]
\sin[\pi\alpha_{X}(t)]}\left(\frac{s}{s_{0}}
\right)^{\overline{\alpha}_X},
\end{eqnarray}
with $\overline{\alpha}_X=\alpha'_{X}(t-m^{2}_{X})$, $\alpha_{X}(t)=
J_{X}+\alpha'_{X}(t-m^{2}_{X})$. Here $\alpha_X$ denotes the Regge trajectory for the meson $X$
as a function of $t=-Q^2$ with the slope $\alpha'_X$, $J_X$ and $m_X $
stand for the spin and mass of the meson, respectively. The phase factors of the propagators are taken as positive unity since they do not affect the results.

With Eqs. (\ref{effective-interactions})-(\ref{Regge}), the contribution from interference of Figs. \ref{ep-scattering}(d) and \ref{Born-photo-production}(a) can be calculated directly. After combining it with the Born contribution, the reduced unpolarized cross section is expressed as
\begin{eqnarray}\label{meson-to-sigma}
\sigma_{un,th}^{1\gamma+2\gamma(M)} &=& \sigma_{un,th}^{1\gamma}+gf_0s^{\overline{\alpha}_T} (G_M (1 + \varepsilon )\tau +
2G_E \varepsilon),
\end{eqnarray}
and the polarized observables $P_t,P_l$ are expressed as
\begin{eqnarray}\label{meson-to-PTPL}
P_{t,th}^{1\gamma+2\gamma(M)} &=& P_{t,th}^{1\gamma}\frac{\sigma_{un,th}^{1\gamma}}{\sigma_{un,th}^{1\gamma+2\gamma(M)}}-\frac{gf_1s^{\overline{\alpha}_T}(G_E + 2G_M)}{\sigma_{un,th}^{1\gamma+2\gamma(M)}},\nonumber \\
P_{l,th}^{1\gamma+2\gamma(M)} &=&  P_{l,th}^{1\gamma}\frac{\sigma_{un,th}^{1\gamma}}{\sigma_{un,th}^{1\gamma+2\gamma(M)}}+\frac{gf_2s^{\overline{\alpha}_T} G_M }{\sigma_{un,th}^{1\gamma+2\gamma(M)}},
\end{eqnarray}
where $f_0=  \sqrt {\tau (1 +
\tau )(1 + \varepsilon ) / (1 - \varepsilon )}, f_1=\tau \sqrt {\varepsilon(1+\varepsilon)(1 + \tau )/2}, f_2=\tau ^{3 /
2}\sqrt {(1 + \tau )} (2\varepsilon + 1),\sqrt{s}$ is the center of mass of the $ep$ system and $g$ is expressed as
\begin{eqnarray}
g=\textrm{Re}[\frac{-4iM_N^{4}g_{Tee,1}g_{Tpp,1}\alpha'_{T}}{\alpha\Gamma[\alpha_{T}(t)-J_{T}+1]
\sin[\pi\alpha_{X}(t)]}\left(\frac{1}{s_{0}}\right)^{\overline{\alpha}_T}]. \nonumber
\end{eqnarray}

The most important property of the above three corrections is that only the $2^{++}$ meson-exchange gives contributions due to the zero mass of the electron. This property lead to the interesting result that the three corrections to $\sigma_{un,th}^{1\gamma}, P_{t,l,th}^{1\gamma}$ are only dependent on one new parameter $g$ which is a constant at fixed $Q^2$. This makes it possible to extract $g$ by fitting the unpolarized experimental data with Eq.(\ref{meson-to-sigma}) and then use such extracted parameters to predict the TPE corrections to $P_{t,l,th}^{1\gamma}$. A nenefit of such extracting and prediction is its universality since we have not assumed any special model dependent calculation for the coupling. If the extracted $g$ is zero then it naturally means the meson-exchange mechanism can be neglected and the extracted $G_{E,M}$ naturally return to those extracted by Eq.(\ref{OPE-sigma}), and if the extracted $g$ is not zero, then it means the meson-exchange effect really exists or there are some other similar notable physical effects beyond the OPE and usual TPE corrections from Fig.\ref{ep-scattering}(a,b).
The second important property of the corrections is that they  all vanish when $\epsilon \rightarrow 1$ due to the factor $s^{\overline{\alpha}_T}$ which is expected by unitarity.

In the practical calculation, we take $\overline{\alpha}_T=0.8(t-1.3^2 $GeV$^2$)\cite{Regge-meson-exchange} and the detailed analysis shows that the results are not sensitive to the slope of $\overline{\alpha}_T$ in the region [0.7,0.9].

To estimate the TPE contributions from Fig.\ref{ep-scattering}(a,b), we use the simple hadronic model and include $N$ and $\Delta$ as the intermediate states. For the TPE contributions from $N$, we take the same parameters as \cite{TPE-hadronic-model}. For the TPE contribution from $\Delta$, we improve the choice of the coupling parameters and form factors of $\Gamma_{\gamma N\Delta}$ used in \cite{TPE-hadronic-model} by taking  $(g_1,g_2,g_3)$=$(6.59,9.06,7.16)$ and
\begin{eqnarray}\label{D3}
F^{(1)}_{\Delta}&=&F^{(2)}_{\Delta}=\left(\frac{\Lambda_{1}^2}{q^2-\Lambda_{1}^{2}}\right)^{2}
\frac{-\Lambda_{3}^2}{q^2-\Lambda_{3}^{2}},\\ \nonumber
F^{(3)}_{\Delta}&=&\left(\frac{\Lambda_{1}^2}{q^2-\Lambda_{1}^{2}}\right)^{2}\frac{-\Lambda_{3}^2}{q^2-\Lambda_{3}^{2}}
\left [a\frac{-\Lambda_{2}^2}{q^2-\Lambda_{2}^{2}}+
(1-a)\frac{-\Lambda_{4}^2}{q^2-\Lambda_{4}^{2}}\right ],\nonumber
\end{eqnarray}
with $\Lambda_{1,2,3,4}=(0.84,2,\sqrt{2},0.2)$GeV and $a=-0.3$. Such coupling parameters and form factors of $\gamma N \Delta$ are much closer to the physical results \cite{SNYang-PR} than those used in \cite{TPE-hadronic-model}.  With these inputs, the contribution from the interference of Figs. \ref{ep-scattering}(a,b) and\ref{Born-photo-production}(a)
can be calculated directly as \cite{TPE-hadronic-model} and the detailed analysis of these two contributions can see \cite{zhouhq2014}.

\section{Numerical results and discussion}

\begin{figure}[htbp]
\center{\epsfxsize 5.0 truein\epsfbox{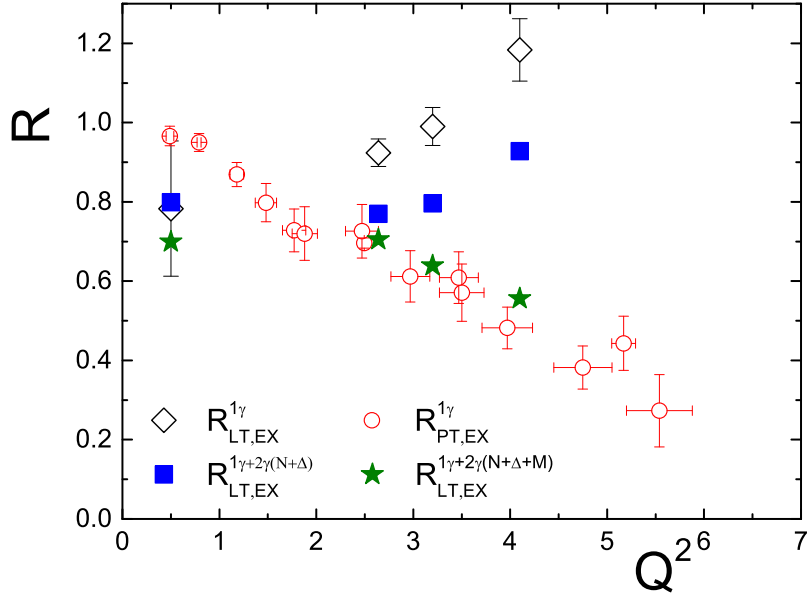}}
\caption{Extracted $R$ by the LT and PT methods. $R_{LT,Ex}^{1\gamma}$ refers to the extracted $R$ by Eq.(\ref{OPE-sigma}) from the experimental data without any TPE corrections , $R_{LT,Ex}^{1\gamma+2\gamma(N+\Delta)}$ and $R_{LT,Ex}^{1\gamma+2\gamma(N+\Delta+M)}$ refer to the extracted $R$ by Eqs. (\ref{OPE-sigma}) and (\ref{meson-to-sigma}) after applying the usual TPE corrections from Fig.\ref{ep-scattering}(a,b) to the experimental data, respectively. The unpolarized experimental data are taken from \cite{Ex-Rosenbluth-2006} and $R_{PT,Ex}^{1\gamma}$ are taken from \cite{Ex-polarized}. The error bar of experimental data is taken as the weight in the fitting. }
\label{RLTPT}
\end{figure}

To show the meson-exchange corrections to the extracted $R$ in the LT method, at first we apply the usual TPE corrections from Figs .\ref{ep-scattering}(a,b) \protect\footnotemark[1] \protect\footnotetext[1]{In this paper, all the TPE correction from $N$ intermediate state refers to the one that the soft part has been deducted  as done in \cite{TPE-hadronic-model}.} to the experimental data sets of unpolarized cross sections as done in \cite{Arrinton2007}, and then extract the corresponding $R$ from the TPE-corrected data using Eqs. (\ref{OPE-sigma}) and (\ref{meson-to-sigma}), respectively. We name such extracted $R$ as $R_{LT,Ex}^{1\gamma+2\gamma(N+\Delta)}$ and $R_{LT,Ex}^{1\gamma+2\gamma(N+\Delta+M)}$, respectively. The results are presented in Fig.\ref{RLTPT} where only the recent precise experimental data \cite{Ex-Rosenbluth-2006} are taken and the error bar of experimental data is taken as the weight in the fitting.

The results in Fig.\ref{RLTPT} clearly show that when no TPE contributions are considered, the extracted $R_{LT,Ex}^{1\gamma}$ \cite{Ex-Rosenbluth-2006} are totally inconsistent with that by the PT method $R_{PT,Ex}^{1\gamma}$ \cite{Ex-polarized}.
After considering the usual TPE contributions from Figs. \ref{ep-scattering}(a,b), the extracted $R_{LT,Ex}^{1\gamma+2\gamma(N+\Delta)}$
are much closer to $R_{PT,Ex}^{1\gamma}$, while an obvious discrepancy still exists for $Q^2=3.2,4.1$ GeV$^2$ cases.
When the meson-exchange contribution is also considered, the extracted $R_{LT,Ex}^{1\gamma+2\gamma(N+\Delta+M)}$ are naturally close to $R_{PT,Ex}^{1\gamma}$.

In the following, we will show that in the region where most of the PT experiment is measured, $R_{PT,Ex}^{1\gamma}$ are close to $R_{PT,Ex}^{1\gamma+2\gamma(N+\Delta+M)}$ with $R_{PT,Ex}^{1\gamma+2\gamma(N+\Delta+M)}$ defined as the extracted $R$ by the PT method after applying the TPE correction to the experimental  PT data. The combination of the above two properties means $R_{LT,Ex}^{1\gamma+2\gamma(N+\Delta+M)}$ are consistent with $R_{PT,Ex}^{1\gamma+2\gamma(N+\Delta+M)}$ and the larger discrepancy of $R$ between the PT and LT methods can be well understood.

\begin{table}[htbp]
\begin{tabular}
{|p{50pt}<{\centering}|p{40pt}<{\centering}|p{40pt}<{\centering}|p{40pt}<{\centering}||p{40pt}<{\centering}|p{40pt}<{\centering}|p{40pt}<{\centering}|}
\hline&
\multicolumn{3}{|p{122pt}<{\centering}||}{results with the error bar as weight in the fitting } &
\multicolumn{3}{p{122pt}<{\centering}|}{results without weight \newline in the fitting}  \\
\hline  $Q^2$(GeV$^2)$ & $G_M$ & $R$ & $g$ & $G_M$ & $R$ & $g$ \\
\hline2.46&0.136&0.704&-0.439&0.136&0.704&-0.461 \\
\hline3.2&0.101&0.639&-1.203&0.101&0.639&-1.213 \\
\hline4.1&0.066&0.556&-6.377&0.067&0.352&-8.590 \\
\hline
\end{tabular}
\caption{Extracted parameters $G_M,R,g$ by Eq.(\ref{meson-to-sigma}) after applying the usual TPE corrections  from Fig.\ref{ep-scattering}(a,b) to experimental data\cite{Ex-Rosenbluth-2006}.}\label{GMGEg}
\end{table}

We list the extracted $G_M, R, g$ by the above method in Tab.\ref{GMGEg}, where, for comparison, the extracted results without any weight are also presented. The comparison shows the extracted results are almost independent on the weight at $Q^2=2.64,3.2$ GeV$^2$, this means the experimental data sets are very precise at these two $Q^2$. From Table \ref{GMGEg}, we can see that the absolute magnitude of $g$ increases when $Q^2$ increases. At first glance, this property seems very un-natural, while actually the coupling $g$ is always accompanied by a factor $s^{\overline{\alpha}_T}$ which decreases very quickly when $Q^2$ increases since $s\geq M_N^2(1+\tau)(1+2\tau+2\sqrt{\tau(1+\tau)})$.

\begin{figure}[htbp]
\center{\epsfxsize 6.4 truein\epsfbox{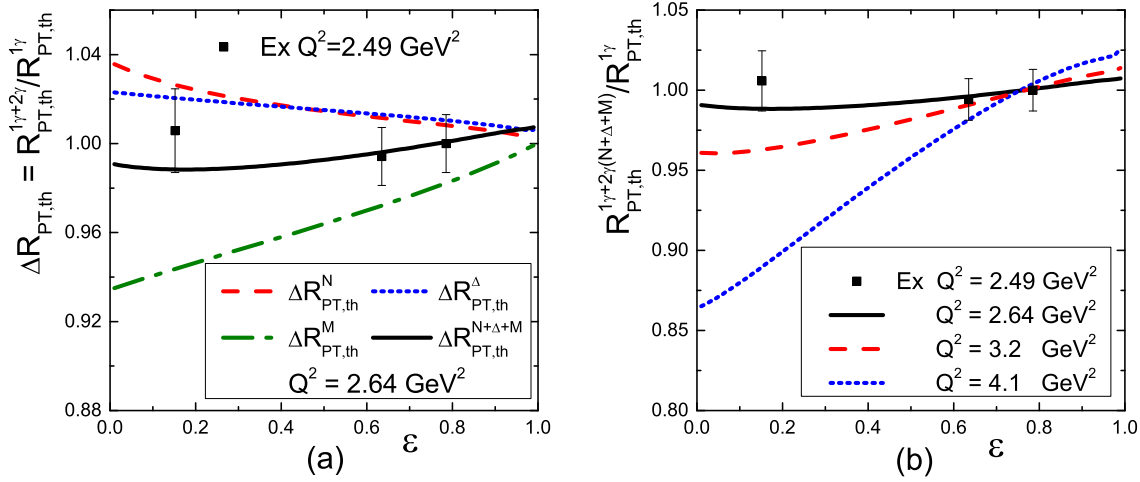}}
\caption{Theoretical estimations of TPE corrections to $R_{PT}$. $\Delta R_{PT,th}^{N,\Delta,M,N+\Delta+M}$ refer to the corresponding theoretical estimations of TPE contributions from $N,\Delta $ intermediate states, meson-exchange and their sum, respectively. The experimental results are taken from \cite{Ex-polarized-Meziane-2011} and normalized at $\epsilon=0.785$.}
\label{RPT}
\end{figure}

\begin{figure}[htbp]
\center{\epsfxsize 6.4 truein\epsfbox{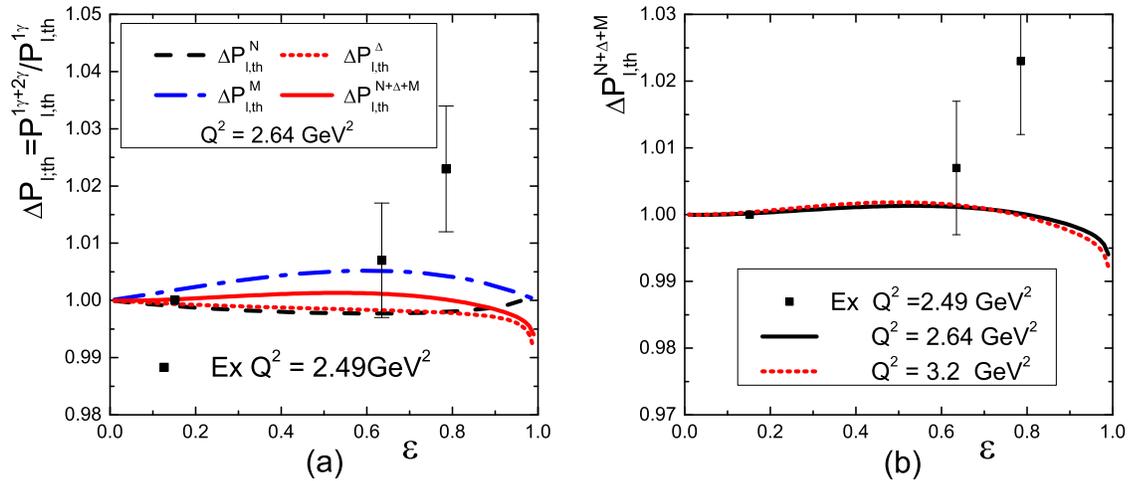}}
\caption{Theoretical estimations of TPE corrections to $P_{l}$. $\Delta P_{l,th}^{N,\Delta,M,N+\Delta+M}$ refer to the theoretical estimations of TPE corrections from $N,\Delta $ intermediate states, meson-exchange and their sum, respectively. The experimental results are taken from \cite{Ex-polarized-Meziane-2011} and normalized at $\epsilon=0.152$.}
\label{PL}
\end{figure}

In the following discussion, we take the $G_M, R, g$ in the left side of Table \ref{GMGEg} as the physical quantities to calculate the polarized observables $P_{t,l,th}^{1\gamma+2\gamma(N,\Delta,M)}$ and their ratio $R_{PT,th}^{1\gamma,1\gamma+2\gamma(N,\Delta,M)}$ which is defined as $-\mu_p\sqrt{\tau(1+\epsilon)/2\epsilon} P_{t,th}^{1\gamma+2\gamma(N,\Delta,M)}/P_{l,th}^{1\gamma+2\gamma(N,\Delta,M)}$, where the indexes $1\gamma$ and $2\gamma(N,\Delta,M)$ refer to the results without and with corresponding TPE contributions, respectively. To compare the  theoretical TPE corrections with the polarized experimental results directly, we define
\begin{eqnarray}
\Delta P_{t,l,th}^{N,\Delta,M} &\equiv& P_{t,l,th}^{1\gamma+2\gamma(N,\Delta,M)}/P_{t,l,th}^{1\gamma}  ,   \nonumber \\
\Delta R_{PT,th}^{N,\Delta,M}&\equiv& R_{PT,th}^{1\gamma+2\gamma(N,\Delta,M)}/R_{PT,th}^{1\gamma}    .
\end{eqnarray}
After all the TPE corrections are included, we expect the following properties if the TPE corrections are the right ones:
\begin{eqnarray}
 P_{t,l,th}^{1\gamma+2\gamma(N+\Delta+M)} &=& P_{t,l,Ex},   \nonumber \\
 R_{PT,th}^{1\gamma+2\gamma(N+\Delta+M)} &=& R_{PT,Ex}^{1\gamma},   \nonumber \\
 R_{PT,th}^{1\gamma}  &=& R_{PT,Ex}^{1\gamma+1\gamma(N+\Delta+M)}  = \mu_p G_E/G_M,
\end{eqnarray}
where $P_{t,l,Ex}$ refer to the measured $P_{t,l}$ by experiment. This results in
\begin{eqnarray}\label{deltaRPl}
\Delta P_{l,th}^{N+\Delta+M} & = & P_{l,th}^{1\gamma+2\gamma(N+\Delta+M)}/P_{l,th}^{1\gamma} \nonumber \\
&=& P_{l,Ex}/P_{l,Ex}^{Born} , \nonumber \\
\Delta R_{PT,th}^{N+\Delta+M} &=&  R_{PT,th}^{1\gamma+2\gamma(N+\Delta+M)}/R_{PT,th}^{1\gamma}  \nonumber \\
&=& R_{PT,Ex}^{1\gamma}/R_{PT,Ex}^{1\gamma+2\gamma(N+\Delta+M)}  \nonumber \\
&\approx& R_{PT,Ex}^{1\gamma}/R_{PT,Ex}^{1\gamma}|_{\epsilon \approx 1} ,
\end{eqnarray}
where the approximate equal is due to the unitarity that TPE corrections to the extracted $R$ by the PT method are assumed to be zero at $\epsilon=1$, and $P_{l,Ex}^{Born}$ is estimated in a corresponding experiment \cite{Ex-polarized-Meziane-2011}. By these relations, we can compare our theoretical results with the experimental data directly. The numerical results are presented in Figs. \ref{RPT} and\ref{PL}.

For the $Q^2=2.64$ GeV$^2$ case, Fig. \ref{RPT}(a) shows  that at small $\epsilon$ the corrections from the usual TPE contributions $\Delta R^{N,\Delta}_{PT,th}$ are large and positive while the corrections from meson-exchange $\Delta R^{M}_{PT,th}$ are large and negative, and they are canceled to some degree which results in the small magnitude of the full TPE corrections $\Delta R^{N+\Delta+M}_{PT,th}$. At large $\epsilon>0.7$ all three corrections are small. For the $Q^2=3.2$ GeV$^2$ case, the situation is similar and the full TPE correction $\Delta R^{N+\Delta+M}_{PT,th}$ shown in Fig. \ref{RPT}(b) are also small for almost all $\epsilon$. For the $Q^2=4.1$ GeV$^2$ case, the comparable experimental $R_{PT,Ex}^{1\gamma}$ at $Q^2=4.0$ GeV$^2$ is measured at $\epsilon=0.71$ \cite{Ex-polarized}, and the corresponding $\Delta R^{N+\Delta+M}_{PT,th}$ is as small as about 3\% in this region.  By Eq. (\ref{deltaRPl}), the smallness of $\Delta R^{N+\Delta+M}_{PT,th}$ means $R_{PT,Ex}^{1\gamma}$ are close to $R_{PT,Ex}^{1\gamma+2\gamma(N+\Delta+M)}$ in the region we discussed, combining with the property that $R_{LT,Ex}^{1\gamma+2\gamma(N+\Delta+M)}$ are close to $R_{PT,Ex}^{1\gamma}$, we get the above conclusion that $R_{LT,Ex}^{1\gamma+2\gamma(N+\Delta+M)}$ are consistent with $R_{PT,Ex}^{1\gamma+2\gamma(N+\Delta+M)}$.

Figure \ref{RPT}(b) also shows the full TPE correction $\Delta R^{N+\Delta+M}_{PT,th}$ decreases when $Q^2$ decreases. The behaviors of $\Delta R^{N+\Delta+M}_{PT,th}$ at $Q^2=2.64,3.2,4.1$ GeV$^2$ strongly suggest it may be close to 1 for almost all $\epsilon$ at $Q^2=2.49$ GeV$^2$ and are consistent with the recent experimental results of $\epsilon$ dependence of $R_{PT,Ex}^{1\gamma}$ \cite{Ex-polarized-Meziane-2011}  which can not be explained by other model dependent calculations such as the simple hadronic model, pQCD, and GDPs method.

Figure \ref{PL} shows that the behavior of $\Delta P_{l,th}^{N+\Delta+M}$ is much closer to the experiment results than  $\Delta P_{l,th}^{N+\Delta}$, while a considerable discrepancy with experimental data still exists at large $\epsilon$. Since the experimental error bars of $P_{l,Ex}$ are not small, it is a little difficult to give a certain conclusion on such a discrepancy at present and further more precise experiments will be a good and interesting test.

\begin{figure}[htbp]
\center{\epsfxsize 5.0 truein\epsfbox{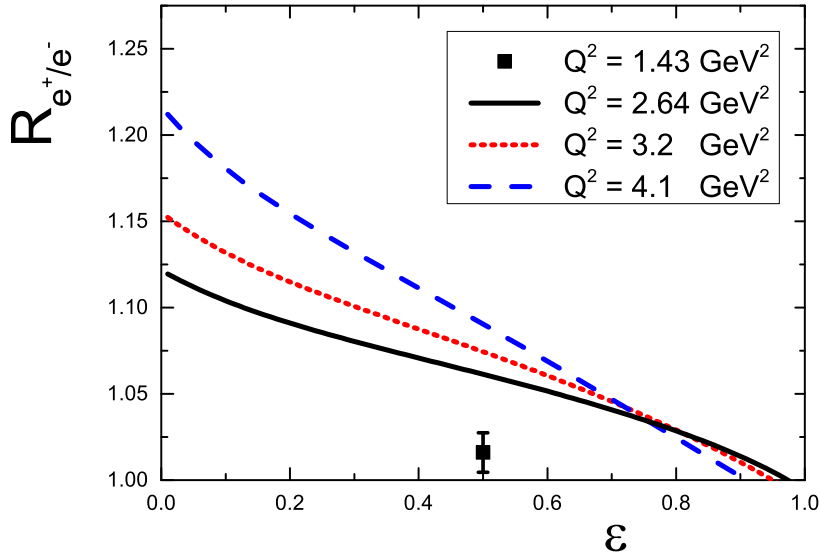}}
\caption{The theoretical estimation of ratio $R_{e^+/e^-}$ at $Q^2=2.64,3.2,4.1$ GeV$^2$ after considering the full TPE corrections from $N,\Delta $ intermediate states and meson-exchange, the experimental data is taken from \cite{Rpm-VEPP}.}
\label{Rpm}
\end{figure}

Using the parameters listed in Table \ref{GMGEg} and including the usual TPE corrections from $N$ and $\Delta$ intermediate states, the ratio $R_{e^+/e^-}\equiv \sigma_{un,e^+p\rightarrow e^+p}/\sigma_{un,,e^-p\rightarrow e^-p}$ can also be calculated directly and the corresponding numerical results are presented in Fig. \ref{Rpm}. The numerical results at $Q^2=2.64,3.2,4.1$ GeV$^2$ show a similar magnitude and properties with that predicted by \cite{Vanderhaeghen-2011-EPJA} where both the unpolarized and polarization data are used for fitting. Comparing with the smallness of $R_{e^+/e^-}$ at $Q^2 < 2$ GeV$^2$ \cite{Rpm-VEPP}, the results suggest the measurement of $R_{e^+e^-}$ at $Q^2=2.5$ GeV$^2$ and small $\epsilon$ will be a good test to the theoretical study of TPE effects.

To summarize, we suggest a new dynamical form of TPE effect in elastic $ep$ scattering and estimate its contributions to extracted $R's$ by the LT and PT methods, $P_l$ and $R_{e^+/e^-}$ with one unknown universal coupling parameter $g$ at fixed $Q^2$. We find after combining such contributions with the usual TPE contributions from box and crossed-box diagrams, the extracted $R's$ by the LT method from the recent precise experimental data \cite{Ex-Rosenbluth-2006} are naturally close to those measured by the PT method. And using the extracted $G_{M}, R$ and $g$ by LT method, the $\epsilon$ dependence of $R$ by the PT method at $Q^2=2.49$ GeV$^2$ \cite{Ex-polarized-Meziane-2011} can be described well, also our results for $R_{e^+e^-}$ are similar with those predicted by \cite{Vanderhaeghen-2011-EPJA}. The full results suggest the meson-exchange mechanism may play an important role in elastic $ep$ scattering and more precise experimental data at $Q^2=2.5$ GeV$^2$ will be a good test.

\section{Acknowledgments}
H.-Q.Z thanks S.N. Yang for helpful discussions. This work is supported by the National Sciences Foundations of China
under Grant No. 11375044 and in part by the Fundamental Research Funds for the Central Universities under Grant No. 2242014R30012.

\end{document}